\documentclass[
    ,final            
  ]
  {aipproc}
  
\usepackage{amsmath}
\usepackage{amssymb}

\layoutstyle{8x11single}

\begin{document}

\title{\hspace{-2.5mm}Minimal Resonant Leptogenesis and Lepton Flavour Violation}

\classification{11.30.Er, 14.60.St, 98.80.Cq}
\keywords      {Resonant Leptogenesis, Lepton Flavour Violation}

\author{Frank F. Deppisch}{
  address={Department of Physics and Astronomy, University College
    London, London WC1E 6BT, United Kingdom} 
}

\author{Apostolos Pilaftsis}{ address={Consortium for Fundamental
    Physics, School of Physics and Astronomy, University of
    Manchester,\\ Manchester M13 9PL, United Kingdom} }

\begin{abstract}
We discuss minimal non-supersymmetric models of resonant leptogenesis,
based  on  an  approximate  flavour symmetries.   As  an  illustrative
example, we consider  a resonant $\tau$-leptogenesis model, compatible
with universal  right-handed neutrino masses  at the GUT  scale, where
the required heavy-neutrino mass splittings are generated radiatively.
In  particular, we  explicitly demonstrate,  how a  minimum  number of
three  heavy  Majorana  neutrinos   is  needed,  in  order  to  obtain
successful   leptogenesis  and   experimentally  testable   rates  for
processes of  lepton flavour violation,  such as $\mu\to  e\gamma$ and
$\mu \to e$ conversion in nuclei.
\end{abstract}

\maketitle


\section{Introduction}
\label{sec:introduction}

The observed baryon asymmetry in  the Universe (BAU), which amounts to
a baryon-to-photon ratio of number densities $\eta_B\approx 6.2 \times
10^{-10}$~\cite{WMAP, Komatsu:2010fb},  provides one of  the strongest
pieces    of    evidence    for    physics   beyond    the    Standard
Model~(SM)~\cite{reviews}. One interesting scenario for explaining the
BAU is  leptogenesis~\cite{FY}. A potentially  interesting alternative
to  GUT-scale  leptogenesis is  the  framework  of low-scale  resonant
leptogenesis (RL)~\cite{APRD, PU}. Within this framework, the lowering
of  the   scale  may   rely  on  a   dynamical  mechanism,   in  which
heavy-neutrino  self-energy  effects~\cite{LiuSegre}  on the  leptonic
asymmetry   become    dominant~\cite{Paschos}   and   get   resonantly
enhanced~\cite{APRD}, when  a pair of  heavy Majorana neutrinos  has a
mass difference comparable to the heavy neutrino decay widths.

Flavour effects due to heavy-neutrino Yukawa couplings~\cite{EMX} play
an important  role in  models of RL~\cite{APtau,PU2}.   In particular,
in~\cite{APtau}   a  scenario   was  put   forward,   called  resonant
$\tau$-leptogenesis~(R$\tau$L),  in which  the BAU  originates  from a
$\tau$-lepton  asymmetry, resonantly produced  by quasi-in-equilibrium
decays of heavy Majorana neutrinos. In a R$\tau$L model, the generated
excess in the $L_\tau$ number will be converted into the observed BAU,
provided  the $L_\tau$-violating  reactions are  not strong  enough to
wash  out such  an  excess. In  such  a scenario,  the heavy  Majorana
neutrinos can  be as light as  100~GeV and have  sizeable couplings to
two  of the  charged  leptons, specifically  to  electrons and  muons.
Consequently,  depending on  the  flavour dynamics  of heavy  neutrino
Yukawa coupling effects, phenomenologically  testable models of RL can
be built that could be probed  at the LHC or in low energy experiments
of lepton  number violation~(LNV) and  lepton flavour violation~(LFV),
such as  $\mu\to e\gamma$ and  neutrinoless double beta decay.   As we
demonstrated  in   \cite{Deppisch:2010fr},  R$\ell$L  models   can  be
constructed with a degenerate heavy  neutrino mass spectrum at the GUT
scale.   In  a minimal  non-supersymmetric  framework, the  splittings
required  for successful  RL are  generated  radiatively~\cite{GJN} by
renormalization group  (RG) evolution  and can therefore  be naturally
comparable to the decay widths of the heavy neutrinos.

\section{$\tau$-Lepton Flavour Symmetry}
\label{sec:model}

The  leptonic Yukawa  and  Majorana sectors  of  the SM  symmetrically
extended  with one  singlet right-handed  neutrino $\nu_{iR}$  per $i$
family (with $i=1,2,3 = e,\mu ,\tau$) are given by the Lagrangian
\begin{equation}
  \label{Lym}
	-\; {\cal L}_{Y,M}\ =\
	 \bar{L} \Phi\,  {\bf h}^\ell\, l_{R}\ +\
	\bar{L}\tilde{\Phi}\, {\bf h}^{\nu}\, \nu_R\ +\ 
         \bar{\nu}_R^C\, {\bf m}_M\, \nu_R\ +\ \text{H.c.},
\end{equation}
where  $\Phi$ is  the SM  Higgs  doublet and  $\tilde{\Phi} =  i\tau_2
\Phi^*$  its  isospin conjugate.   Moreover,  we  have suppressed  the
generation index  $i$   from  the  left-handed  doublets  $L_i
=(\nu_{iL}, l_{iL})^T$, the  right-handed charged leptons $l_{iR}$ and
the right-handed  neutrinos $\nu_{iR}$, while  ordinary multiplication
between vectors and  matrices is implied.

To obtain  a phenomenologically relevant model in  this minimal setup,
at  least  3 singlet  heavy  Majorana  neutrinos $\nu_{1,2,3\,R}$  are
needed and  these have to  be nearly degenerate  in mass~\cite{APtau}.
To ensure  the latter,  we assume that  to leading order,  the singlet
Majorana sector is SO(3) symmetric, i.e.
\begin{equation}
  \label{MSSO3}   
	{\bf m}_M\ =\ m_N {\bf 1}_3\ +\ {\Delta \bf m}_M\; , 
\end{equation}
where ${\bf 1}_3$ is the  $3\times 3$ identity matrix and $\Delta {\bf
  m}_M$ is a  general SO(3)-breaking matrix induced by  RG effects. As
we will discuss below,  compatibility with the observed light neutrino
masses  and   mixings  requires  that   $({\Delta  \bf  m}_M)_{ij}/m_N
\stackrel{<}{{}_\sim   }  10^{-7}$,   for   electroweak-mass  Majorana
neutrinos,  i.e.~for  $m_N \approx  0.1$--1~TeV.   We will  explicitly
demonstrate, how such an SO(3)-breaking matrix ${ \Delta \bf m}_M$, of
the required order, can be  generated radiatively via the RG evolution
of  the right-handed  neutrino mass  matrix ${\bf  m}_M$ from  the GUT
scale $M_X \approx 2\times 10^{16}~{\rm GeV}$ to the mass scale of the
right-handed neutrinos $m_N$.

In the physical charged-lepton  mass basis, the SO(3) symmetry imposed
on the singlet Majorana sector  at the GUT scale $M_X$ gets explicitly
broken by a set of neutrino Yukawa couplings to the subgroup of lepton
symmetries:  ${\rm U(1)}_{L_e+L_\mu}\times {\rm  U(1)}_{L_\tau}$.  The
flavour  charge assignments  that give  rise  to such  a breaking  are
presented in Table~\ref{tab:ChargesRtauL}.

\begin{table}[t]
\centering
\begin{tabular}{lccccc}
\hline
        & $L_e, e_R$ & $L_\mu, \mu_R$ & $L_\tau, \tau_R$ 
        & $\nu_1$ & $\nu_2\pm i\nu_3$    \\
\hline
$U(1)_{L_e+L_\mu}$ & +1 & +1 &  0 & 0 & $\pm 1$ \\
$U(1)_{L_\tau}$    &  0 &  0 & +1 & 0 &      0  \\
\hline
\end{tabular}
\caption{\it Flavour charge assignments for the breaking ${\rm SO(3)}\to
  {\rm U(1)}_{L_e+L_\mu}\times {\rm U(1)}_{L_\tau}$.}\label{tab:ChargesRtauL}  
\end{table}

As   a   consequence  of   the   ${\rm  U(1)}_{L_e+L_\mu}\times   {\rm
  U(1)}_{L_\tau}$ symmetry, the  neutrino Yukawa coupling matrix takes
on the general form:
\begin{equation}
  \label{hnutau}
	{\bf h}^\nu\ =\
	\left(\begin{array}{ccc}
		0  & a e^{-i\pi/4}  & a e^{i\pi/4} \\
		0  & b e^{-i\pi/4}  & b e^{i\pi/4} \\
		0  & 0                & 0 
	\end{array}\right)\ 
	+
	\left(\begin{array}{ccc}
		\epsilon_e    & 0 & 0 \\
		\epsilon_\mu  & 0 & 0 \\
		\epsilon_\tau & \kappa_1 e^{-i(\pi/4-\gamma_1)} & 
		\kappa_2 e^{i(\pi/4-\gamma_2)}
	\end{array}\right),
\end{equation}
where the  second term vanishes, if  the symmetry is  exact. In this
symmetric limit, the light neutrinos  remain massless to all orders in
perturbation theory, whilst $a$ and $b$ are free unconstrained complex
parameters. The phases accompanying these parameters in~(\ref{hnutau})
are simply chosen for convenience  to maximize the lepton asymmetry in
leptogenesis when  $a$ and $b$ are  real.  In order to  give masses to
the light neutrinos, the second  term describes the departure from the
flavour  symmetric limit  with  $|\epsilon_{e,\mu,\tau}|, \kappa_{1,2}
\ll |a|, |b|$ and the phases $\gamma_{1,2}$ are unrestricted.

It is important  to notice that the flavour  structure of the neutrino
Yukawa couplings ${\bf h}^\nu$  is preserved through the RG evolution,
as long as the flavour symmetry  is only weakly broken.  In detail, RG
effects violate  the SO(3)-invariant form of the  Majorana mass matrix
${\bf m}_M(M_X)  = m_N  {\bf 1}_3$ by  the 3-by-3 matrix  ${\Delta \bf
  m}_M$.  In the  leading-log approximation, the SO(3)-breaking matrix
${\Delta \bf m}_M$ receives the following radiative corrections:
\begin{equation}
  \label{eq:mNFirstOrder}
	{\Delta \bf m}_M =
	- \frac{m_N}{8\pi^2}\ln\left(\frac{M_X}{m_N}\right)
	\text{Re}\left[ {\bf h}^{\nu\dagger}(M_X) {\bf h}^\nu(M_X)
          \right].
\end{equation}
Given the  form invariance of ${\bf  h}^\nu$ under RG  effects, we may
therefore  define all  input parameters  at the  right-handed neutrino
mass scale $m_N$,  where the matching with the  light neutrino data is
performed.

We   may  now  determine   the  Yukawa   parameters  $(a,b,\epsilon_e,
\epsilon_\mu,  \epsilon_\tau)$, in  terms of  the  light-neutrino mass
matrix  ${\bf m}^\nu$  in  the positive  and  diagonal charged  lepton
Yukawa basis. To do so, we first notice that the chosen symmetry ${\rm
  U(1)}_{L_e+L_\mu}\times  {\rm U(1)}_{L_\tau}$ ensures  the vanishing
of  the  light  neutrino  mass  matrix ${\bf  m}^\nu$.   However,  the
symmetry-breaking  parameters induce  a non-zero  light  neutrino mass
matrix ${\bf  m}^\nu$, which to  leading order in these  parameters is
given by~\cite{PU2,Deppisch:2010fr}
\begin{eqnarray}
  \label{mnutree}
	{\bf m}^\nu &=&
        -\ \frac{v^2}{2}\, {\bf h}^\nu\, {\bf m}_M^{-1} {\bf h}^{\nu
          {\rm T}}\  =\ \frac{v^2}{2m_N}\, 
	\bigg(\frac{{\bf h}^\nu {\Delta \bf m}_M {\bf h}^{\nu {\rm T}}}{m_N}\
		-\ {\bf h}^\nu {\bf h}^{\nu {\rm T}}
	\bigg) \nonumber\\
	&=&
	-\ \frac{v^2}{2m_N}
	\begin{pmatrix}
		\frac{\Delta m_N}{m_N}a^2-\epsilon_e^2  & 
		\frac{\Delta m_N}{m_N}ab-\epsilon_e\epsilon_\mu & 
		-\epsilon_e\epsilon_\tau \\
    	\frac{\Delta m_N}{m_N}ab-\epsilon_e\epsilon_\mu & 
		\frac{\Delta m_N}{m_N}b^2-\epsilon_\mu^2          &
		-\epsilon_\mu\epsilon_\tau \\
		-\epsilon_e\epsilon_\tau & 
		-\epsilon_\mu\epsilon_\tau &
		-\epsilon_\tau^2
	\end{pmatrix},
\end{eqnarray}
where $v = 2M_W/g_w = 245$~GeV  is the vacuum expectation value of the
SM    Higgs   field~$\Phi$.    In    deriving   the    last   equation
in~(\ref{mnutree}),  we have  also  assumed that  $\sqrt{\frac{\Delta
    m_N}{m_N}}\kappa_{1,2}\ll  \epsilon_{e,\mu,\tau}$,  where  $\Delta
m_N$ stands for the expression
\begin{equation}
  \label{DmN}
	\Delta m_N \equiv
   2({\Delta \bf m}_M)_{23}\: +\: i\Big[ ({\Delta \bf m}_M)_{33}
- ({\Delta \bf m}_ M)_{22} \Big]
 =  -\ \frac{m_N}{8\pi^2}\ln\bigg(\frac{M_X}{m_N}\bigg)
	\Big[ 2\kappa_1\kappa_2\sin(\gamma_1+\gamma_2) 
		+ i(\kappa_2^2-\kappa_1^2) \Big]\; .
\end{equation}
As a consequence of the flavour symmetry ${\rm U(1)}_{L_e+L_\mu}\times
{\rm     U(1)}_{L_\tau}$,     the    symmetry-violating     parameters
$\epsilon_{e,\mu,\tau}$ and $\kappa_{1,2}$  enter the tree-level light
neutrino    mass    matrix    ${\bf   m}^\nu$    in    (\ref{mnutree})
quadratically. This  in turn implies that  for electroweak-scale heavy
neutrinos $m_N  \sim v$,  the symmetry-breaking Yukawa  couplings need
not  be much  smaller  than  the electron  Yukawa  coupling $h_e  \sim
10^{-6}$.   Moreover,  one  should  observe  that  only  a  particular
combination of SO(3)-violating terms $({\Delta \bf m}_M)_{ij}$ appears
in   ${\bf   m}^\nu$   through   $\Delta  m_N$.    Nevertheless,   for
electroweak-scale  heavy  neutrinos  with  mass  differences  $|\Delta
m_N|/m_N \stackrel{<}{{}_\sim}  10^{-7}$, one should  have $|a|,\, |b|
\stackrel{<}{{}_\sim}  10^{-2}$  to  avoid  getting  too  large  light
neutrino masses much above~0.5~eV. Finally, we note that one-loop $Z$-
and  Higgs-boson effects  can induce  additional contributions  to the
light  neutrino mass  matrix  ${\bf m}^\nu$~\cite{AZPC}.   As well  as
being loop suppressed,  these radiative contributions are proportional
to the symmetry-breaking parameters and can therefore be neglected for
the R$\tau$L scenario under study~\cite{PU2}.

Given  the analytic  form~(\ref{mnutree}) of  the light  neutrino mass
matrix,  we  may  directly   compute  the  neutrino  Yukawa  couplings
$(a,b,\epsilon_e,  \epsilon_\mu,   \epsilon_\tau)$,  as  functions  of
$m_N$, the  phenomenologically constrained neutrino  mass matrix ${\bf
  m}^\nu$  and  the  symmetry-breaking parameters  $\kappa_{1,2}$  and
$\gamma_{1,2}$:
\begin{eqnarray}
  \label{eq:a2}
	a^2 &=& 
	\frac{2m_N}{v^2}\ \frac{8\pi^2}{\ln(M_X/m_N)}\
	\bigg( m^\nu_{11} - \frac{(m^\nu_{13})^2}{m^\nu_{33}}\bigg)\
	\Big[ 2\kappa_1\kappa_2\sin(\gamma_1+\gamma_2) 
		+ i(\kappa_2^2-\kappa_1^2)\Big]^{-1}\; , \nonumber\\
	b^2 &=& 
	\frac{2m_N}{v^2}\ \frac{8\pi^2}{\ln(M_X/m_N)}\
	\bigg( m^\nu_{22} - \frac{(m^\nu_{23})^2}{m^\nu_{33}}\bigg)\
	\Big[ 2\kappa_1\kappa_2\sin(\gamma_1+\gamma_2) 
		+ i(\kappa_2^2-\kappa_1^2)\Big]^{-1}\; ,\\
	\epsilon_e^2 &=& 
	\frac{2m_N}{v^2}\ \frac{(m^\nu_{13})^2}{m^\nu_{33}}\; ,\qquad
	\epsilon_\mu^2\ =\ 
	\frac{2m_N}{v^2}\ \frac{(m^\nu_{23})^2}{m^\nu_{33}}\; , \qquad
	\epsilon_\tau^2\ =\ 
	\frac{2m_N}{v^2}\ m^\nu_{33}\; .\nonumber
\end{eqnarray}
Since  the approximate  light-neutrino  mass matrix  ${\bf m}^\nu$  in
(\ref{mnutree})  has rank  2,  the lightest  neutrino mass  eigenstate
$\nu_1$ will be massless in this approximation. The relations given in
(\ref{eq:a2}) will be  used to obtain numerical estimates  of the BAU,
in terms of $m_N$  and the symmetry-breaking parameters $\kappa_{1,2}$
and $\gamma_{1,2}$,  for both normal and  inverted hierarchy scenarios
of light neutrinos.

\section{Resonant Leptogenesis}
\label{sec:leptogenesis}

Within  the  framework  of   leptogenesis,  a  net  non-zero  leptonic
asymmetry results  from the CP-violating decays of  the heavy Majorana
neutrinos $N_\alpha$ into the  left-handed charged leptons $l^-_L$ and
light  neutrinos $\nu_{lL}$.  Consequently, we  have to  calculate the
partial decay width  of the heavy Majorana neutrino  $N_\alpha$ into a
particular lepton flavour $l$,
\begin{equation}
\label{GammaN}
	\Gamma_{\alpha l}\ =\
	\Gamma(N_\alpha \to l^-_L + W^+)\: +\: 
	\Gamma(N_\alpha\to \nu_{lL} + Z, H)\; . 
\end{equation}

In   RL  models,  resumming   the  absorptive   parts  of   the  heavy
Majorana-neutrino self-energy transitions $N_\beta \to N_\alpha$ plays
an   important   role    in   the   computation   of   $\Gamma_{\alpha
  l}$~\cite{APRD,PU}.

The  leptonic asymmetries for  each individual  lepton flavour  can be
expressed  in   terms  of  the  resummed   neutrino  Yukawa  couplings
$\overline{\bf h}^\nu_{l\alpha}$~\cite{PU,Pilaftsis:2008qt} as
\begin{equation}
  \label{deltaN}
	\delta_{\alpha l}\ \equiv\ 
	\frac{ 
		\Gamma_{\alpha l}^{\phantom{C}}\: -\:
		\Gamma_{\alpha l}^C 
	}{ 
		\sum\limits_{l = e,\mu ,\tau}
		\Big(\Gamma _{\alpha l}^{\phantom{C}}\: +\:
		\Gamma_{\alpha l}^C\Big)
	}\ =\
	\frac{\big|\overline{\bf h}^\nu_{l\alpha}\big|^2\: -\:
			\big|\overline{\bf h}^{\nu C}_{l\alpha}\big|^2}
	{ 
	\big(\overline{\bf h}^{\nu \dagger}\, 
	\overline{\bf h}^{\nu \phantom{\dagger}}\!\!\big)_{\alpha\alpha}\: +\: 
	\big(\overline{\bf h}^{\nu C\dagger}\, 
	\overline{\bf h}^{\nu C\phantom{\dagger}}\!\!\big)_{\alpha\alpha}
	}\ .
\end{equation}

The analytic results for  the leptonic asymmetries $\delta_{\alpha l}$
simplify    considerably    in    the    2-heavy    neutrino    mixing
limit~\cite{APRD,PU},
\begin{equation}
\label{dCPla}
\delta_{\alpha l} \ \approx\ \frac{{\rm Im} 
\big[\, ({\bf h}^{\nu\dagger}_{\alpha l}{\bf h}^{\nu\phantom{\dagger}}_{l\beta})\,
({\bf h}^{\nu\dagger}_{\phantom{l}}{\bf
    h}^{\nu\phantom{\dagger}}_{\phantom{l}}\!\!)_{\alpha\beta}\big]} 
{({\bf     h}^{\nu\dagger}_{\phantom{l}}
{\bf h}^{\nu\phantom{\dagger}}_{\phantom{l}}\!\!)_{\alpha\alpha}\, ({\bf
    h}^{\nu\dagger}_{\phantom{l}}
{\bf h}^{\nu\phantom{\dagger}}_{\phantom{l}}\!\!)_{\beta\beta}}\
\frac{(m^2_{N_\alpha} - m^2_{N_\beta})\, m_{N_\alpha}\,
  \Gamma^{(0)}_{N_\beta}}{ (m^2_{N_\alpha} - m^2_{N_\beta})^2\: +\: 
m^2_{N_\alpha} \Gamma^{(0)2}_{N_\beta}}\ ,
\end{equation}
where  $\alpha,  \beta  =   1,2$  and  $\Gamma^{(0)}_{N_a}$  is  the
tree-level decay width of  the heavy Majorana neutrino~$N_\alpha$. The
following  two  conditions  for  having resonantly  enhanced  leptonic
asymmetries   $\delta_{\alpha   l}    \sim   {\cal   O}(1)$   may   be
derived~\cite{APRD,PU2}:
\begin{eqnarray}
  \label{ResCond}
\mbox{(i)}&& |m_{N_\alpha} -
  m_{N_\beta}|\  \sim\ \frac{\Gamma_{N_{\alpha,\beta}}}{2}\ ,\\
\mbox{(ii)}&& \frac{\big| {\rm Im} 
\big[\, ({\bf h}^{\nu\dagger}_{\alpha l}{\bf h}^{\nu\phantom{\dagger}}_{l\beta})\,
({\bf h}^{\nu\dagger}_{\phantom{l}}{\bf
    h}^{\nu\phantom{\dagger}}_{\phantom{l}}\!\!)_{\alpha\beta}\big]\big|} 
{({\bf     h}^{\nu\dagger}_{\phantom{l}}
{\bf h}^{\nu\phantom{\dagger}}_{\phantom{l}}\!\!)_{\alpha\alpha}\, ({\bf
    h}^{\nu\dagger}_{\phantom{l}}
{\bf
  h}^{\nu\phantom{\dagger}}_{\phantom{l}}\!\!)_{\beta\beta}}\ \sim\ 1\ .
\end{eqnarray}
Within  our RL  scenarios, the  first condition  in~(\ref{ResCond}) is
naturally  fulfilled   as  the  heavy-neutrino   mass  splittings  are
generated via  RG effects and are  of the required  order.  The second
condition is  crucial as  well and controls  the size of  the leptonic
asymmetries.   A detailed  comparison of  the predictions  obtained by
different    approaches   to    resonant    leptogenesis   is    given
in~\cite{Deppisch:2010fr}.

The  Boltzmann   equations  for   the  heavy  Majorana   neutrino  and
lepton-number  densities, $\eta_{N_{1,2,3}}$, $\eta_{L_{e,\mu,\tau}}$,
normalized to the photon number density $n_\gamma$ are given
by~\cite{PU2,Deppisch:2010fr} 
\begin{eqnarray}
  \label{BEN}
\frac{d\eta_{N_\alpha}}{dz} & = &
	-\ \frac{z}{n_\gamma H_N}\; 
	\bigg( \frac{\eta_{N_\alpha}}{\eta^{\rm eq}_N} - 1 \bigg)\,
	\gamma^{N_\alpha}_{L\Phi}\; , \\[2mm]
  \label{BEDL}
\frac{d\eta_{L_l}}{dz} & = &
	\frac{z}{n_\gamma H_N}\
     \bigg[\; \sum_{\alpha=1}^3
   \bigg(\frac{\eta_{N_\alpha}}{\eta^{\rm eq}_N} - 1\bigg)\, 
        \delta_{\alpha l}\gamma^{N_\alpha}_{L\Phi}\ -\ 
  \frac{2}{3}\; \eta_{L_l}\!\sum_{k=e,\mu,\tau} \bigg(
			\gamma^{L_l\Phi}_{L^C_k\Phi^\dagger} +
			\gamma^{L_l\Phi}_{L_k\Phi} \bigg)\nonumber\\
&& -\ \frac{2}{3}\sum_{k=e,\mu,\tau}\, \eta_{L_k}
 \bigg(	\gamma^{L_k\Phi}_{L^C_l\Phi^\dagger} -
 \gamma^{L_k\Phi}_{L_l\Phi} \bigg)\; \bigg]\; ,
\end{eqnarray}
where  $\alpha =  1,\,2,\,3$ and  $l =  e,\,\mu,\,\tau$.  In addition,
$H_N\approx  17\times  m_N^2/M_{\rm P}$  is  the  Hubble parameter  at
$T=m_N$,  where $M_{\rm  P}  = 1.2\times  10^{19}$~GeV  is the  Planck
mass.  The $T$-dependence of  the BEs~(\ref{BEN})  and~(\ref{BEDL}) is
expressed in terms of the dimensionless parameter $z = \frac{m_N}{T}$,
and $\eta_N^{\rm eq}$  is the equilibrium number density  of the heavy
Majorana  neutrino $N_\alpha$,  normalized  to the  number density  of
photons.  The BEs~(\ref{BEN}) and  (\ref{BEDL}) include  the collision
terms $\gamma(X \to Y) \equiv \gamma^X_Y$ for the decays $N_\alpha \to
L_l\Phi$, as well as  the $\Delta L=0,2$ resonant scattering processes
$L_k\Phi \to L_l\Phi$ and $L_k\Phi \to L^C_l \Phi^\dagger$~\cite{PU2}.

\begin{figure}[t]
\centering
\includegraphics[clip,width=0.65\textwidth]{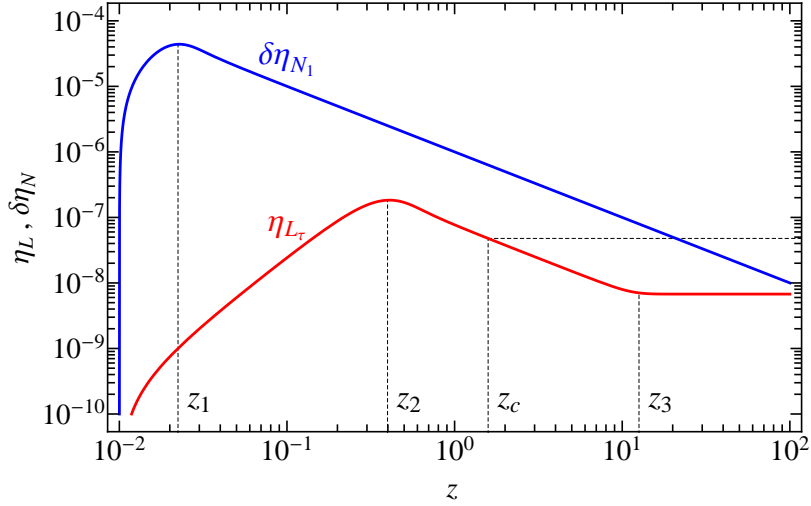}
\caption{Numerical solutions to the Boltzmann equations for
  $\delta\eta_{N_1} = \eta_{N_1}/\eta_N^{\rm eq} - 1$ and
  $\eta_{L_\tau}$, respectively, in a R$\tau$L model with $m_N =
  220$~GeV, $K_{N_1} = \Gamma_{N_1}/H_N = 10^6$, $K_\tau=10^2$,
  $\delta_\tau=10^{-6}$.}
\label{fig:eta_z} 
\end{figure}
The   lepton   asymmetries    are   then   partially   converted   via
$(B+L)$-violating sphaleron processes~\cite{KRS}. For temperatures $T$
larger  than the  critical temperature  $T_c \approx  135$~GeV  of the
electroweak   phase   transition,   the   conversion  of   the   total
lepton-to-photon   ratio  $\sum_{l=e,\mu,\tau}   \eta_{L_l}$   to  the
baryon-to-photon  ratio $\eta_B^c$ at  $T_c$ is  given by  $\eta_B^c =
-\frac{28}{51} \sum_{l=e,\mu,\tau} \eta_{L_l}$. After dilution of this
baryon asymmetry $\eta_B^c$ by  standard photon interactions until the
recombination epoch,  $\eta_B \approx 1/27  \eta_B^c$, the theoretical
prediction can  now be compared  with the current  observational value
for the baryon-to-photon asymmetry~\cite{Komatsu:2010fb}: $\eta_B^{\rm
  obs} = \left(6.20\pm 0.15\right) \times 10^{-10}$.

In the limit of strong  washout, the baryon asymmetry $\eta_B$ becomes
relatively  independent  of the  initial  values  of $\eta_{L_l}$  and
$\eta_{N_\alpha}$. In this case, the Boltzmann equations can be solved
analytically   and   the   resulting   BAU   can   be   estimated   to
be~\cite{Deppisch:2010fr}
\begin{equation}
	\eta_B \approx
	\frac{-3\cdot 10^{-2}\delta_\tau}
	{K_\tau {\rm min}\Big[m_N/T_c, 1.25\ln(25K_\tau)\Big]},
\end{equation}
with   the   total  $\tau$   asymmetry   $\delta_\tau  =   \sum_\alpha
\delta_{\alpha\tau}$  and the  total $\tau$  asymmetry  washout factor
$K_\tau  =   \sum_\alpha  \Gamma_{N_\alpha\to  L_\tau\phi}/H_N$.  This
formula  provides a  good approximation  of the  generated BAU  in the
R$\tau$L scenario  within the regime  $K_\tau \stackrel{>}{{}_\sim} 5$
for  a  right-handed  neutrino mass  scale  of  the  order of  the  EW
scale.  Hence,  to account  for  the  observed  BAU, a  $\tau$  lepton
asymmetry $\delta_\tau \stackrel{>}{{}_\sim} 10^{-7}$ is required.

\section{Lepton Flavour Violation and $\theta_{13}$}
\label{sec:results}

Heavy Majorana-neutrino loop effects may induce sizeable LFV couplings
to the  photon and  the $Z$  boson. These couplings  give rise  to LFV
decays, such  as $\mu \to e\gamma$~\cite{CL},  $\mu \to eee$~\cite{IP}
and $\mu \to e$ conversion in  nuclei. In the R$\tau$L model, only two
of the  right-handed neutrinos have appreciable  $e$- and $\mu$-Yukawa
couplings    $a,b\approx    10^{-2}$    for   $\kappa_{1,2}    \approx
10^{-5}-10^{-3}$. The  LFV branching ratio $B(\mu\to  e\gamma)$ can be
expressed as~\cite{Deppisch:2010fr}
\begin{equation}
\label{eq:BllgammaApprox}
	B(\mu\to e\gamma)\ \approx 
	8.0\cdot 10^{-4}\times  
	g\left(\frac{m_N}{m_W}\right)
	\frac{v^4}{m_N^4}a^2 b^2,
\end{equation}
with the loop function $g(x)$, possessing the limits $g\to 1$ for $m_N
\gg m_W$  and $g\to 1/16$  for $m_N=m_W$. This  theoretical prediction
can be favourably contrasted with the current experimental upper limit
$B(\mu\to e\gamma) < 2.4\cdot 10^{-12}$~\cite{Adam:2011ch}.

\begin{figure}[t]
\centering
\includegraphics[clip,width=0.45\textwidth]{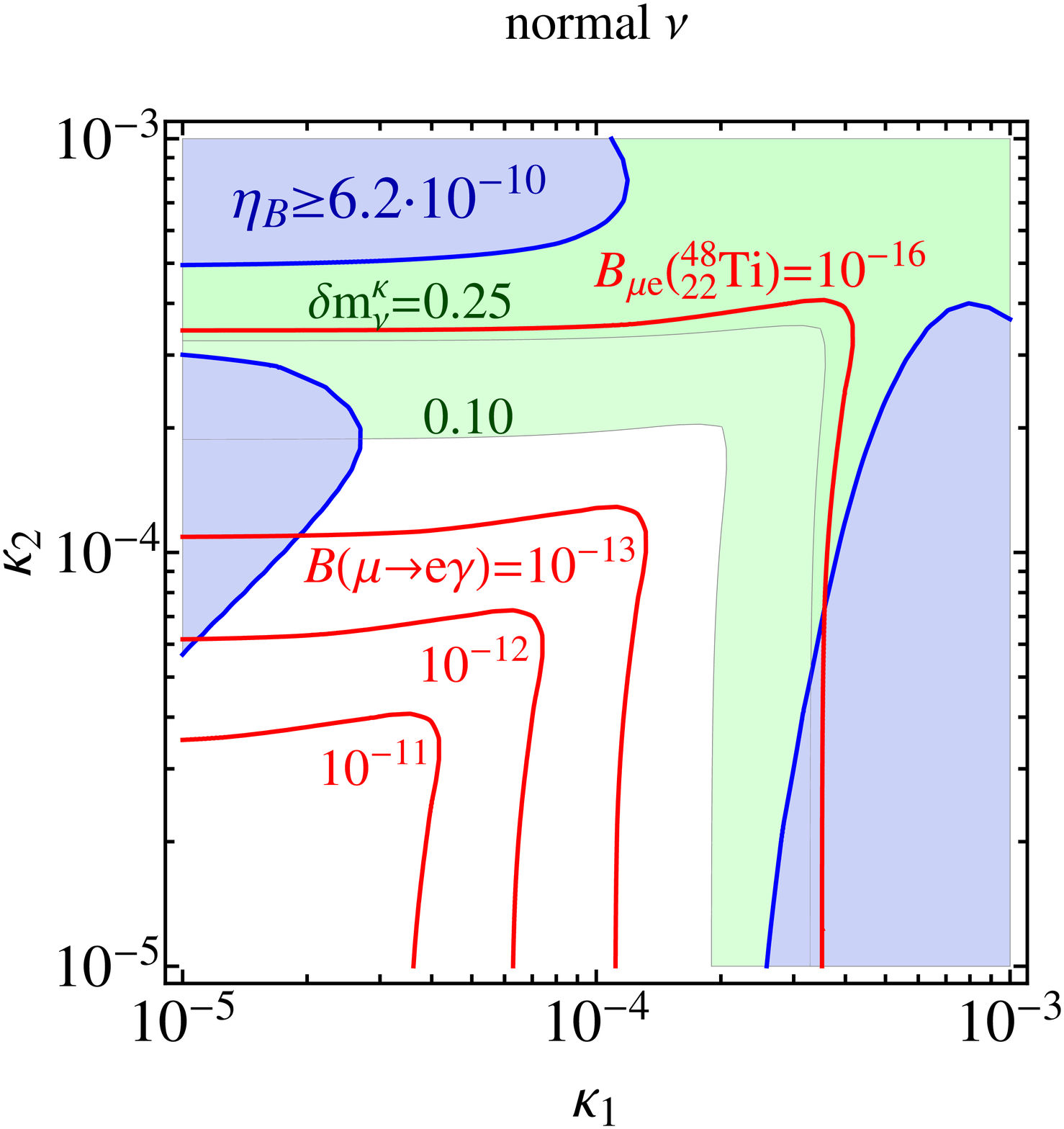}
\includegraphics[clip,width=0.45\textwidth]{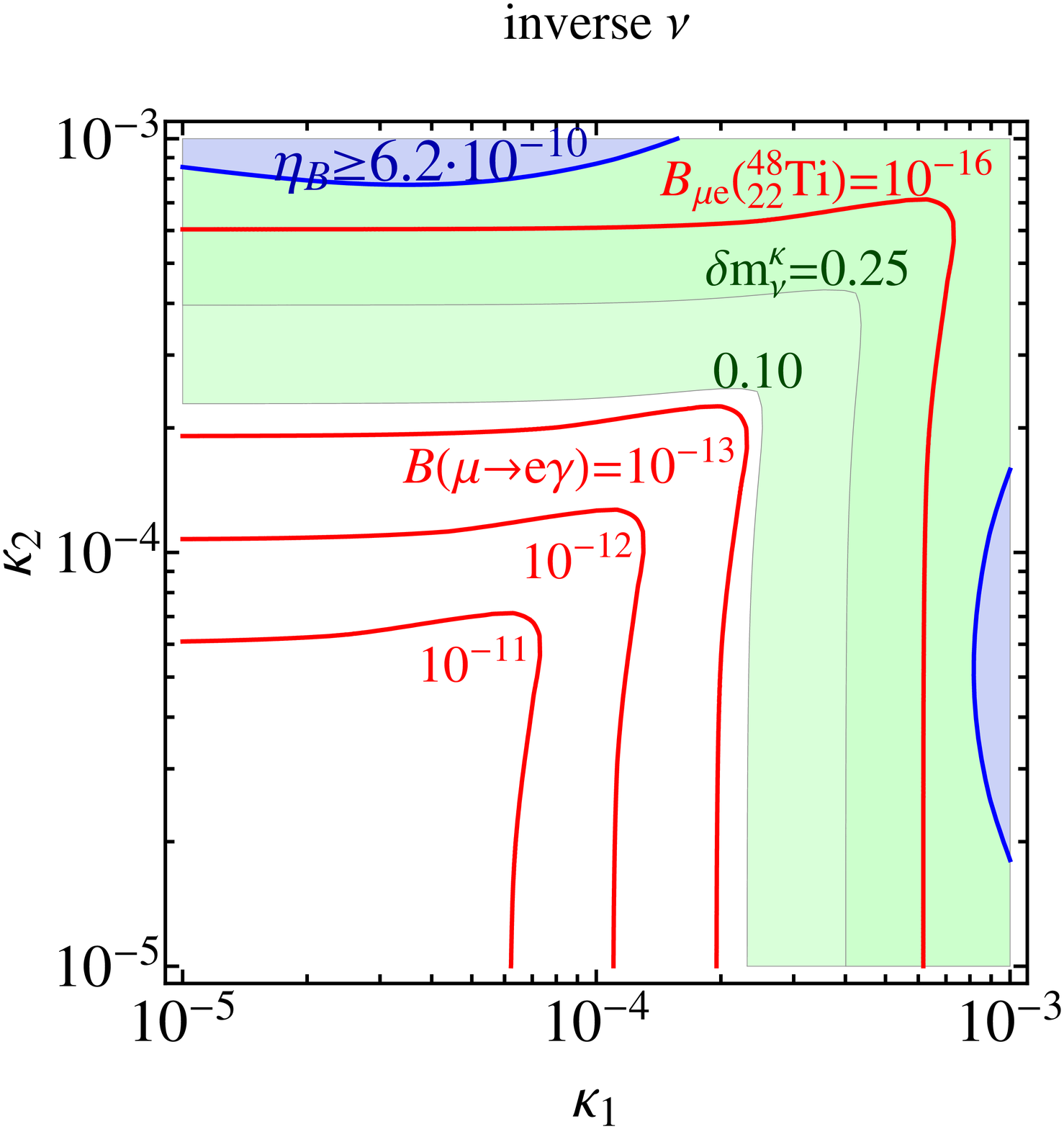}
\caption{Baryon asymmetry $\eta_B$ (blue contours) and the LFV
  observables $B(\mu \to e\gamma)$ and $B_{\mu e}({}^{48}_{22}{\rm
    Ti})$ (red contours) as functions of $\kappa_1$ and $\kappa_2$ in
  the R$\tau$L model with $m_N=120$~GeV, assuming a normal (left) and
  inverse (right) light neutrino mass spectrum. The remaining
  parameters are: $\gamma_1=3\pi/8$, $\gamma_2=\pi/2$, $\phi_1=\pi$,
  $\phi_2=0$, $\text{Re}(a)>0$. The neutrino oscillation parameters
  are set at their best fit values, with $\sin^2 \theta_{13} =
  0.027$. The blue shaded regions denote the parameter space where the
  baryon asymmetry is larger than the observed value. The green shaded
  areas labeled as `$\,\delta {\rm m}_\nu^\kappa=0.25$' and `$\,0.10$'
  indicate the parameter space where the inversion of the
  light-neutrino mass matrix is violated at the 25\% and 10\% level,
  respectively.}
\label{fig:k1_k2_RtauL_75_100} 
\end{figure}

As  discussed above,  the neutrino  oscillation  data can  be used  to
determine  the  theoretical  parameters   of  the  R$\tau$L  model  by
inverting  the seesaw  formula  and solving  for  the neutrino  Yukawa
couplings  $a$,  $b$  and  $\epsilon_{e,\mu,\tau}$. In  addition,  the
electroweak-scale flavour structure  of the right-handed neutrino mass
matrix~${\bf   m}_M$    is   generated   from    a   flavour-universal
heavy-neutrino mass matrix ${\bf m}_M(M_X) = m_N {\bf 1}_3$ at the GUT
scale $M_X$, after taking into  account RG-running effects. We use the
best  fit values  of~\cite{Tortola:2012te} for  the  measured neutrino
oscillation parameters, and  the three unknown CP phases  of the light
neutrino  sector are  taken  into account  by  scanning over  possible
combinations of the extreme CP parities.

Fig.~\ref{fig:k1_k2_RtauL_75_100} shows the numerical estimates of the
baryon asymmetry $\eta_B$ and the LFV observables $B(\mu \to e\gamma)$
and   $B_{\mu   e}({}^{48}_{22}{\rm  Ti})$,   as   functions  of   the
Yukawa-coupling  parameters $\kappa_1$ and  $\kappa_2$, in  a R$\tau$L
model with $m_N=120$~GeV and with either a normal or an inverted light
neutrino mass  spectrum. The  phases of the  $\kappa_{1,2}$ parameters
are: $\gamma_1=3\pi/8$,  $\gamma_2=1/2\pi$.  This choice approximately
enhances  the overlap of  the areas  with successful  baryogenesis and
large LFV  rates.  The  blue shaded areas  denote the  parameter space
where  the numerically predicted  baryon asymmetry~$\eta_B$  is larger
than   the   observational  value   $\eta^{\rm   obs}_B  =   6.2\times
10^{-10}$. These  areas should be regarded as  representing regions of
viable parameter space for  successful leptogenesis, given the freedom
of   re-adjusting  the   CP  phases   $\gamma_{1,2}$  of   the  Yukawa
couplings~$\kappa_{1,2}$. The  green areas denote  the parameter space
where the  assumed approximation $\kappa_{1,2}\ll a,b$  is violated at
10\% or 25\%.

\begin{figure}[t]
\centering
\includegraphics[clip,width=0.7\textwidth]{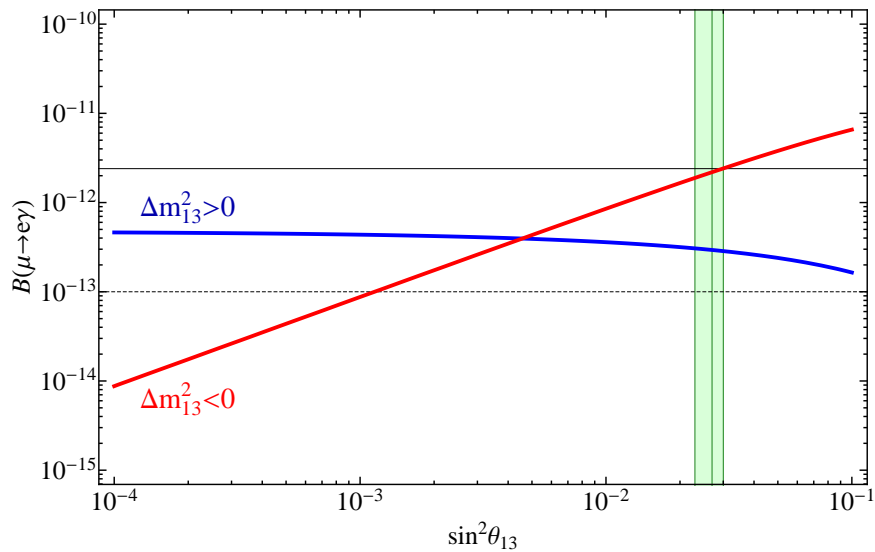}
\caption{Branching  ratio   $B(\mu\to  e\gamma)$  as   a  function  of
  $\sin^2\theta_{13}$, in R$\tau$L scenarios  with a normal (blue) and
  inverted (red) light neutrino mass spectra. The model parameters for
  these  two cases are  as in  Fig.~\ref{fig:k1_k2_RtauL_75_100}, with
  $\kappa_1=10^{-5}$  and  $\kappa_2=10^{-4}$.   The horizontal  solid
  (dashed)  line  denotes  the  current  (expected  future)  limit  on
  $B(\mu\to e\gamma)$.  The vertical  green band indicates the allowed
  range of  $\sin^2\theta_{13}$, with an  uncertainty of $\pm1\sigma$;
  the  middle vertical line  shows the  current experimental  best fit
  value for $\sin^2\theta_{13}$~\cite{Tortola:2012te}.}
\label{fig:theta13} 
\end{figure}

The branching ratio $B(\mu \to e\gamma)$ is enhanced in the case of an
inverted   hierarchical   light   neutrino  mass   spectrum   ($\Delta
m^2_{13}<0$).   Instead,  in  the normal  hierarchical  light-neutrino
scenario with $\Delta  m^2_{13}>0$, $B(\mu\to e\gamma)$ is essentially
independent   of   $\sin\theta_{13}$.    This   is   demonstrated   in
Fig.~\ref{fig:theta13}, showing  $B(\mu\to e\gamma)$ as  a function of
$\sin^2\theta_{13}$,        where       $\kappa_{1}=10^{-5}$       and
$\kappa_2=10^{-4}$. The same figure  shows the allowed area (displayed
by         a          vertical         green         band)         for
$\sin^2\theta_{13}$~\cite{Tortola:2012te},  as  deduced  from the  two
recent   experiments   conducted   by    the   Daya   Bay   and   RENO
collaborations~\cite{Daya,RENO}.  Unlike  in the normal light-neutrino
scenario, the predicted value  for~$\eta_B$ in the R$\tau$L model with
an inverted  hierarchical light neutrino mass spectrum  falls short of
explaining  the BAU  by two  orders  of magnitude,  for a  potentially
observable branching ratio of $B(\mu \to e\gamma) \sim 10^{-13}$.

\section{Conclusions}
\label{sec:conclusions}

Minimal  low-scale seesaw  scenarios  of resonant  leptogenesis are  a
class  of  models with  interesting  implications  for observables  of
charged LFV, such  as $\mu \to e\gamma$ and $\mu  \to e$ conversion in
nuclei. Here we have discussed  aspects of the R$\tau$L realization of
resonant  leptogenesis,  where the  observed  BAU  originates from  an
individual  $\tau$  lepton-number   asymmetry  which  gets  resonantly
enhanced via the out-of-equilibrium  decays of nearly degenerate heavy
Majorana  neutrinos.  The   required  mass  splittings  are  generated
radiatively via RG effects with universal right-handed neutrino masses
at the GUT scale.

Analogous  to the  R$\tau$L  case discussed  here,  R$e$L and  R$\mu$L
scenarios  with   approximately  protected  $e$,   $\mu$  asymmetries,
respectively,   may   be  constructed~\cite{Deppisch:2010fr}.    Heavy
Majorana  neutrinos in  such R$\ell$L  scenarios  can be  as light  as
100~GeV, whilst their  couplings to two of the  charged leptons may be
large, so as  to lead to LFV  effects that could be tested  by the MEG
and the COMET/PRISM  experiments. In the R$\tau$L model  with a normal
light  neutrino mass hierarchy,  there is  a sizeable  model parameter
space with  successful leptogenesis and large LFV  process rates, with
$B(\mu\to  e\gamma)\approx  10^{-12}$.   This  prediction  is  largely
independent  of  $\sin^2\theta_{13}$  and  the  other  light  neutrino
oscillation parameters. On the other  hand, in the R$\tau$L model with
inversely  hierarchical  light  neutrinos,  $B(\mu  \to  e\gamma)$  is
linearly proportional  to $\sin^2\theta_{13}$, and can  be enhanced by
more than one order of magnitude compared to the normal hierarchy case
for  $\sin^2\theta_{13}$  close to  its  upper experimental  $2\sigma$
limit.   Unfortunately,  the  generated baryon  asymmetry~$\eta_B$  is
suppressed in  this scenario, and  to test the viable  parameter space
for successful  leptogenesis would require an experiment  for $\mu \to
e$  conversion  in nuclei  which  is  sensitive  to $B_{\mu  e}\approx
10^{-17}$--$10^{-16}$.


\begin{theacknowledgments}
AP thanks the organizers of  the GUT2012 workshop in Kyoto, Japan, for
their  kind  hospitality.   He  also gratefully  acknowledges  partial
support   by  the   Lancaster--Manchester--Sheffield   Consortium  for
Fundamental Physics, under STFC research grant: ST/J000418/1.
\end{theacknowledgments}

\bibliographystyle{aipproc}

\end{document}